\documentclass[a4paper]{jpconf}
\usepackage{amssymb}
\usepackage{amsmath}
\usepackage{graphicx}

\def\etal{et al. }
\def\Journal#1#2#3#4{{#1} {\bf #2}, #3 (#4)}

\def\CMP{\em Commun. Math. Phys.}

\def\GRC{\em Grav.Cosmol}

\def\ITP{\em Int. J. Theor. Phys.}

\def\JHE{\em J. High Ener. Phys.}

\def\LRR{\em Living Rev.Rel.}

\def\MPL{{\em Mod. Phys. Lett.} A}

\def\NPB{{\em Nucl. Phys.} B}

\def\PLB{{\em Phys. Lett.}  B}

\def\PRD{{\em Phys. Rev.} D}
\def\PRL{\em Phys. Rev. Lett.}

\def\be{\begin{equation}}
\def\ee{\end{equation}}
\def\bea{\begin{eqnarray}}
\def\eea{\end{eqnarray}}
\def\bes{\begin{equation*}}
\def\ees{\end{equation*}}
\def\beas{\begin{eqnarray*}}
\def\eeas{\end{eqnarray*}}

\begin{document}
\title{And what if gravity is intrinsically quantic ?}

\author{Houri Ziaeepour}

\address{Mullard Space Science Laboratory (MSSL)-University College London, 
Holmbury St.Mary, Dorking RH5 6NT, Surrey, UK}

\ead{hz@mssl.ucl.ac.uk}

\begin{abstract}
Since the early days of search for a quantum theory of gravity the attempts 
have been mostly concentrated on the quantization of an otherwise classical 
system. The two most contentious candidate theories of gravity, sting theory 
and quantum loop gravity are based on a quantum field theory - the latter is 
a quantum field theory of connections on a $SU(2)$ group manifold and former 
a quantum field theory in two dimensional spaces. Here we argue that there is 
a very close relation between quantum mechanics and gravity. Without 
gravity quantum mechanics becomes ambiguous. We consider this observation 
as the evidence for an intrinsic relation between these fundamental laws of 
nature. We suggest a quantum role and definition for gravity in the context of 
a quantum universe, and present a preliminary formulation for gravity in a 
system with a finite number of particles.
\end{abstract}

\section{Introduction and motivations}
Since the beginning of search for quantum theory of gravity all approaches 
have considered the quantization of a classical field theory. Nonetheless the 
idea of an intrinsic relation between gravity and quantum mechanics is not 
new. The first evidence can be claimed to be the black hole entropy and its 
analogy with thermodynamics. However, it is well known that without 
considering Hawking radiation~\cite{hawkingrad} and Bekenstein entropy 
limit~\cite{entropylimit}, this analogy seems to be just a mathematical 
similarity. In fact the black hole entropy is a purely geometrical property 
related to the diffeomorphism symmetry of and independent of the special case 
of the Einstein gravity Lagrangian~\cite{waldnoether,waldpapers}. The 
discovery of similarity between Hawking 
radiation temperature and the equivalent temperature obtained from black hole 
entropy gave another dimension to this as a possible 
mediator between classical and quantum gravity theories. More recently it has 
been proved that in D-brane models~\cite{dbrane} - a subclass of 
compactified string models - at low energies and weak couplings the direct 
counting of the states of an extremal or near extremal black hole lead to an 
entropy similar to the classical black holes~\cite{stringbhentropy}. These 
results enforce the relation between the classical black hole entropy and 
the quantum nature of gravity.

On the way to a quantum theory or a quantum connection for gravity various 
strategies have been taken. Apart from investigations of candidate models 
such as string theory, loop quantum gravity, causal set models, etc., in which 
the metric $g_{\mu\nu}$ or connection $\Gamma^{\rho}_{\mu\nu}$ are quantum 
fields, many authors tried to derive a gravity theory from a field theory in 
a curved space time. For instance T. Padmanabahan~\cite{gremergerindler} has 
used Rindler metric - a frame with constant acceleration - as background 
rather than 
Minkovski metric and a somehow general Lagrangian for gravity. It includes a 
non-trivial surface term on space-like 3-surfaces. By postulating the relation 
between entropy, temperature, and surface gravity obtained from Einstein 
gravity, a dynamical equation similar to one from Einstein gravity has 
been obtained. This exercise shows explicitly that there is a close relation 
between the laws of black hole thermodynamics and Einstein gravity. However, 
black hole entropy is the Noether charge in the models with diffeomorphism 
symmetry~\cite{waldnoether}. Therefore it is not a surprise that in a given 
background the assumption of entropy relation obtained from Einstein gravity 
leads to Einstein equation. This is analogue to gauge symmetry of 
electrodynamics in which the presence of a global conserved charge (surface 
gravity) and the application of Gauss theorem permits to define a 2-form 
(connection) on a 3-surface. Then by using the gauge symmetry (diffeomorphism) 
and its conserved current, one can obtain Maxwell equation and its 
corresponding Lagrangian.

Another popular approach is the construction of a class of theories generally 
called {\it emergent gravity models} in which gravity is considered to be a 
collective low energy/classical effect of a fundamental microscopic model in 
a flat Euclidean or Lorentzian background and very different from classical 
gravity~\cite{gravityemerge}. In this case, gravity is considered to be a 
classical effect similar to condensation process in condense matter physics. 
A curved space time with pseudo-Riemannian metric and diffeomorphism 
invariance have been found in a subset of these models notably models called 
{\it analogue gravity}~\cite{analoguegr}. For the time being, in none of these 
models a tensor field theory with properties similar to what we know from 
Einstein gravity has been found. But there have been progresses in this 
direction. For instance, recently in a work by F. Girelli, 
\etal~\cite{emergenordstrom} a Nordstr\"om scalar gravity model has been 
constructed from a N-scalar field theory on an Euclidean background. 

In all the models mentioned above the 
background space is classical and a quantum field theory is defined on this 
space. The same line of thinking has been considered also in noncommutative 
field theory models with a quantum spacetime~\cite{noncommut}. The main 
physical motivation for introducing noncommutative geometries is the fact 
that due to Heisenberg uncertainty, a very precise measurement of spacetime 
will create very large uncertainty in energy-momentum that can produce a 
black hole preventing further investigation of smaller distances. To solve 
this problem coordinates are considered to have a non-zero commutation. New 
progresses~\cite{noncommutemerge,noncommutmatrix} in this class of 
models show that a classical gravity model similar to Einstein gravity can 
emerge from these models under some conditions and simplifications. As the 
effect of a quantum spacetime in these models is measurable only at Planck 
scales, at present there is no direct way to verify them.

Here we want to consider the relation between classical gravity and quantum 
mechanics in a more intuitive way and see what physical properties join them 
and what make them inconsistent with each other. 

There are a number of unsolved issues in gravity and in quantum mechanics 
that have fundamental consequences for understanding these phenomena. 
Followings are some of them:
\begin {enumerate}
\item Why is gravity a universal force ?
\item Why the Plank Constant $\hbar$ is universal ?
\item Why there is no fundamental mass/length scale in QM ?
\end {enumerate}
The first question above is another way of asking the origin of the 
Equivalence Principle. This issue has been studied intensively in the 
framework of classical and quantum gravity models, and in grand unified 
theories. 
In fact in many candidate models of quantum gravity such as string theory 
Equivalence Principle is a low energy effect that breaks at high energies. 
Nonetheless, at present the upper limit on the breaking of equivalence 
strongly constrains some of low energy quantum gravity models such as brane 
models inspired from D-brane compactification of strings. This rises the issue 
of what is the most fundamental principle in gravity ? Its dependence only on 
energy-momentum or its geometrical origin as the metric of a pseudo-Riemannian 
space, or both. Breaking of Equivalence Principle violates the first 
condition. Therefore if we consider this principle as part of the 
{\it definition} of what we call gravity, models violating it can not be 
considered as a genuine quantum gravity model.

The second universality means that the amount of 
{\it uncertainty or randomness} in all physical systems is the same regardless 
of their mass, size, and couplings. Let's assume that the value of $\hbar$ for 
a particle of mass $m$ is different by a factor of $\alpha$. The 
Schr\"odinger/Klein-Gordon equation becomes:
\bes
(\alpha^2\hbar^2 \Box - m^2) |\psi\rangle = 0 \Longleftrightarrow (\hbar^2 
\Box - m'^2) |\psi\rangle = 0 \quad m' = \frac{m}{\alpha} \label{schrod}
\ees
Therefore the non-universality of $\hbar$ can be removed by redefinition of 
mass. But mass is the gravity charge ! In the same way a different coupling 
$\mathcal G$ to gravity can be removed by redefinition of mass which in its 
turn modifies Schr\"odinger equation. Note that if mass is generated by 
interaction and is related to the Vacuum Expectation Value (VEV) of a field 
this scaling can be performed on the VEV. This operation is allowed in quantum 
mechanics because the energy reference is arbitrary. But considering gravity, 
scaling modifies the coupling of fields to gravity, except in conformal 
models where the scaling can be absorbed in the space volume. In fact it has 
been shown that Einstein and higher order gravity models are related to each 
others by conformal transformation if these models have a scalar field in 
their matter sector~\cite{grconformal}. This is the case for all gauge models 
where the mass of particles are due to VEV of a higgs type scalar fields. In 
this case additional terms from non-conformal matter can be considered as 
non-minimal interaction terms.

The origin of the close dependence of quantum mechanics on gravity is the 
third item above: QM lacks any fundamental energy or length 
scale\footnote{Evidently any energy or length dependent quantity can be used 
as a scale. For instance $\Lambda_{QCD}$ the scale which separates 
perturbative 
and non-perturbative QCD can be also considered as a natural energy scale. 
However, we do not know if this point is unique or cover a range of energies 
with $\alpha_{QCD} \approx 1$. Moreover, for the time being the only 
fundamental interaction with a dimensional coupling constant is gravity, and 
therefore this constant presents the only natural energy scale in physics.}. 
Therefore a theory as fundamental as QM, its extension QFT, and thereby 
Standard Model (SM) of particle physics need the presence of gravity for being 
a complete and applicable theory to Nature. On the other hand, gravity is 
supposed to be just 
an interaction and the general tendency is to see it with the same eye as the 
other interactions (except in emergent gravity models mentioned above). 
Quantum mechanics does not need any of other known interaction e.g. 
electromagnetism to be meaningful. Thus we can conclude that gravity is 
fundamentally different and must have a much closer connection to quantum 
mechanics. 

Although emergent gravity models seem to have shown this connection, gravity 
at low energies rests classical and its connection to a quantum universe 
appears only at Planck scale. Nonetheless, with simple examples we can show 
that even at low energies the non-deterministic nature of quantum mechanics is 
in conflict with a classical (deterministic) universal force like gravity.

\section{Ambiguity of a classical universal force in a quantic universe}
Consider an empty Minkovski space. We are only interested in low energy 
scales, therefore we neglect vacuum fluctuations and all the phenomena 
related to quantum field theory. We add to this space a single particle of 
mass $m$. To remove the ambiguity due to black hole formation, we can 
consider a small but finite size for the particle larger than its 
Schwarzschild radius. In this case if the resolution of the measurement is 
much larger than the size of the particle it can be considered as point-like 
and the argument below is correct up to this approximation. 

As there is no other observer/particle in this space, without loss of 
generality we can consider that the particle is at rest. Because 
we know the energy and momentum of this particle, according to the Heisenberg 
uncertainty rule we lose all information about its location, i.e. at any 
instant of time the particle can be at any point in the space. This has a 
number of consequences: 
\begin{enumerate}
\item A classical particle in a flat Minkovski space breaks the translation 
symmetry to a spherical symmetry and changes the flat metric to Schwarzschild.
\item A quantum particle as defined above does not break the translation 
symmetry. The wave function has the form of a free wave 
$\psi (x) \propto exp (i p_\mu x^\mu)$ and the probability of detecting the 
particle in any finite volume approaches to zero (up to coarse-grain 
approximation mentioned before). To estimate the modification 
of the spacetime due to the presence of this particle, we can use the 
semi-classical Einstein equation:
\be
R_{\mu\nu} -\frac{1}{2} g_{\mu\nu} R = \langle\psi|\hat{T}_{\mu\nu}|\psi
\rangle = \eta_{\mu\nu} \Lambda \quad , \quad \Lambda \rightarrow 0 
\label {desitter}
\ee
The solution of (\ref{desitter}) is simply a De Sitter metric which is 
conformally flat and as $\Lambda = m/V \rightarrow 0$, where $V$ is the 
coarse-grained volume, the metric becomes very close to the 
metric of the initial flat Minkovski space. Note that up to coarse-grain 
approximation this result is independent of the mass of the particle. 
Conclusion: In the context of quantum mechanics a single massive particle does 
not change the spacetime by its gravity !
\item If we add a second particle with only gravitational interaction with 
the first one, we can use Wheeler-DeWitt equation and the above metric as 
background to determine the wave function of the second particle. 
As the background metric is at most De-Sitter, otherwise flat, the wave 
function would be again a free wave but the wave number gradually 
decreases due to the expansion of the space time, and at coarse-grain limit it 
approaches its value in a flat spacetime. Moreover, the homogeneity 
of the spacetime does not change by the presence of a second particle either. 
This means that they do not interact gravitationally. In the same way we can 
add any number of particles and up to coarse-grain limit they do not feel 
each others ! This paradoxical result is evidently inconsistent with the 
behaviour we know from classical gravity.
\item At present we do not have any experimental evidence that tell us what 
happens in such a setup.
\end{enumerate}
The origin of the conflict between classical and quantum predictions is 
the fact that gravity and its dynamical equation are local, in contrast to 
quantum mechanics which is intrinsically nonlocal. Note that a noncommutative 
spacetime and emergent gravity can not solve these ambiguities. The setup of 
this example satisfies the noncommutative relation simply because the 
location of the particle is completely uncertain in all directions. This sort 
of ambiguities should be added to the well known conflict between 
diffeomorphism invariance that leads to a null Hamiltonian and the lack of 
the concept of time evolution in quantum gravity~\cite{curvqm}.

These observations rise the idea that maybe we should think about gravity or 
whatever shows itself as gravity at low energies in another way, presumably 
in a quantic manner. Here we present some ideas about what can be the role of 
gravity with properties we know in a quantum world, and how possibly a 
corresponding field theory can be constructed.

\section{The role of gravity in a quantic Universe}
Consider the phase space of a classical system and the Hilbert space of 
the same system when it is quantized. We can define quantization as the 
definition of an equivalence class in the phase space:
\be 
{\mathcal H} \equiv \Phi / {\mathcal S} \label{hilb}
\ee
where $\Phi$ is the phase space and ${\mathcal S}$ is the set of surfaces with 
area = $\hbar$. For simplicity we neglect the spacetime indexes and 
consider the phase space like a 2-dimensional surface. If we have a system 
consisting of 2 non-interacting particles, their phase space or Hilbert space 
is the direct product of their respective single particle spaces\footnote{Note 
that for simplifying the notification, we interchange the meaning of Hilbert 
space as the vector space to which all the state of a physical system belong 
and the orbit of the Lagrangian operator of the system for a given initial 
condition. The meaning should be clear from the context.}:
\be
\Phi = \Phi_1 \otimes \Phi_2 \quad , \quad {\mathcal H} = 
{\mathcal H}_1 \otimes {\mathcal H}_2 \label{equivclass}
\ee
This means that particles don't {\it feel} each other. However, the assumption 
of no interaction between two particles is not realistic because we know that 
gravity is a general force and whatever the nature of 2 particles, they 
interact through their gravity and therefore:
\be
{\mathcal H} \neq {\mathcal H}_1 \otimes {\mathcal H}_2 \quad ,\quad 
{\mathcal H} = {\mathcal H}_1 \otimes {\mathcal H}_2 
\xrightarrow{\mbox{\footnotesize gravity}} 
{\mathcal H}_1 \otimes {\mathcal H}_2 \label{proj}
\ee
Here we want to suggest a new definition for gravity in a quantum context:\\

{\bf Gravity between quantum particles presents the minimum deviation of 
their Hilbert space from direct products of single-particle Hilbert spaces}\\

This definition is consistent with the universality of gravity and with 
semi-classical results for one particle - a single particle with known energy 
and momentum {\it does not} feel its own gravity ! A direct consequence of 
this property is that such a particle does not make a black hole !
\footnote{At first sight there is an ambiguity in this argument because in 
both classical and quantum mechanics a particle can be composite. We assume 
that equation (\ref{proj}) is applied to fundamental particles. In addition, 
in the frame work of this model the concept of a single particle can be even 
meaningless. See the end of this section for more details.} By 
contrast, if the same particle is localized, then its energy-momentum would be 
completely uncertain similar to particles inside a black hole, which for a far 
observer can be at any place inside the horizon\footnote{Note that the 
Schwarzschild radius here is determined by the energy-momentum uncertainty 
and not by the invariant mass of the particle.}.

The presence of gravity becomes visible only if we have at least 2 particles 
in the system. A nontrivial projection of ${\mathcal H}_1 \otimes 
{\mathcal H}_2$ to itself defines the new Hilbert space ${\mathcal H}$. In 
fact when an interaction is present such a projection happens also in the 
phase space for both classical and quantum systems. What is more important 
in this model and different from others is that gravity mixes the equivalent 
classes:
\be
{\mathcal H} = \frac{\Phi}{\mathcal S} \quad , \quad {\mathcal S} \neq 
{\mathcal S}_1 \otimes {\mathcal S}_2 \label{modequiv}
\ee
This means that uncertainty surfaces for the two particles are not anymore 
separate but define an inseparable subspace of the total phase space $\Phi$. 
This can solve the puzzle of universal $\hbar$. Assuming that this projection 
depends on $G_N$ or equivalently $M_P$, this provides the missing mass/length 
scale for quantum mechanics. The extension of this construction to 
multi-particle systems is trivial and we do not present details here.

The nontrivial mixing of equivalence spaces presenting the Heisenberg 
uncertainty relation means that the uncertainties of coordinate and momentum 
of two particles are not any more independent. This is somehow similar to 
noncommutative models. However, in the latter case the spacetime is considered 
as an independent entity with noncommutative coordinates. Here the nontrivial 
commutation relation is not an intrinsic property but induced. It is possible 
that for large particle number the two models have some sort of relation. We 
leave the investigation of this eventuality to future works.

The nontrivial projection of the combined Hilbert space 
${\mathcal H}_1 \otimes {\mathcal H}_2$ can be considered as a basis 
transformation with the condition that in the new coordinates the transformed 
particles can be considered as free and separable. This basis would play the 
role of a local inertia frame in the classical general relativity and should 
constrain the projection. However, at this level of the development of the 
model we can not prove that it always exists. Therefore we consider that its 
existence is granted. 

We expect that at classical limit the dynamic equation of this model 
approaches Wheeler-DeWitt equation with minimal interaction to gravity:
\be
(\hbar^2 \Box + \frac{R}{6} - m^2)|\psi\rangle = 0  \label{dewittwheeler}
\ee
The general structure of the operator applied to the wave function is:
\be
\hat{P}^\mu \hat{P}_\mu + f_\mu (x)\hat{P}^\mu + h (x)) \quad , \quad 
\hat{P}_\mu \equiv \frac {\partial}{\partial x^\mu} \label{opdewitt}
\ee
We consider the following transformation from one-particle coordinates to 
{\it free} coordinates $X$:
\be
\hat{X}_i = \hat{x}_i \quad , \quad \hat{P}_i = \hat{p}_i + \sum_{j \neq i} 
f (\hat{x}_j, \hat{p}_j)  \label{deffree}
\ee
where $f (\hat{x}_j, \hat{p}_j)$ is an arbitrary function. It is easy to show 
that commutations become:
\bea
&& [\hat{x}_i , \hat{x}_j] = [\hat{X}_i , \hat{X}_j] = 0 \quad i,j = 1, 2, 
\ldots \label{xcommu} \\
&& [\hat{p}_i , \hat{p}_j] = [\hat{P}_i , \hat{P}_j] = 0 \label{pcommu} \\
&& [\hat{x}_i , \hat{p}_i] = -i\hbar \label {xpicommu} \\
&& [\hat{X}_i , \hat{P}_i] = -i\hbar \label {xppicommu} \\
&& [\hat{x}_i , \hat{p}_j] = [\hat{X}_i , \hat{P}_j] = 0 \quad i \neq j 
\label {xpijcommu} \\
&& [\hat{x}_i , \hat{P}_j] = [\hat{x}_i, f (\hat{x}_i , \hat{p}_i)] 
\label {xppijcommu}
\eea
where $f (\hat{x}_i , \hat{p}_i)$ is an arbitrary function. Note that in 
(\ref{xcommu}) to (\ref{xppijcommu}) indexes indicate particles 
and spacetime indexes are neglected. Assuming that the function 
$f (\hat{x}_i, \hat{p}_i)$ is analytical, we can expand it as a polynomial:
\be
f (\hat{x}_i, \hat{p}_i) = \sum_{m,n} \frac{C_{mn} :\hat{x}_i^m \hat{p}_i^n:}
{\Lambda^{n-m-1}} \label {fexpan}
\ee
where $C_{mn}$ is a dimensionless C-number constant and $\Lambda$ is an 
energy scale, presumably a scale comparable to Planck energy. The symbol $::$
indicates that operators $\hat{x}_i$ and $\hat{p}_i$ are ordered such that 
all position operators precede momentum operators. We assume that 
$n-m-1$ the power of $\Lambda$ is always positive such that this term becomes 
important only at energies close to Planck energy. In this case the dominant 
term in (\ref{fexpan}) is the term with $m = -1$ and $n = 1$. If we neglect 
terms with higher power of $\Lambda$ and include the constant $C_{-1,1}$ in 
$\Lambda$:
\bea
&& \hat{P}_i = \hat{p}_i + \sum_{j \neq i} :\frac{\hat{p}_j}{\Lambda 
\hat{x}_j}: \label{ppi} \\
&& [\hat{x}_i , \hat{P}_j] = [\hat{X}_i , \hat{P}_j] = -\frac{i\hbar}
{\Lambda \hat{x}_i} \label {xppfcommu}
\eea
Finally, we apply (\ref{ppi}) to the Schr\"odinger/Klein-Gordon equation 
(\ref{schrod}):
\be
\biggl\{\sum_i \hat{p}_i^2 + 2 \sum_{j \neq i} :\frac{\hat{p}_i\hat{p}_j}
{\Lambda \hat{x}_j}: + \sum_{j,k \neq i} :\frac{\hat{p}_j\hat{p}_k}{\Lambda^2 
\hat{x}_j\hat{x}_k}: + \sum_i m_i^2 \biggr \}|\psi\rangle = 0 \label{schrodgravity}
\ee
The operator part of equation (\ref{schrodgravity}) has the same structure as 
(\ref{opdewitt}) except that it contains discrete operators 
functioning on different particles. We leave the extension of this formalism 
to a continuum for another work.

The choice of commutation relations and the definition of the function $f$ 
are not unique. For instance, similar to noncommutative models one can assume 
that $\hat{x}_i$ and $\hat{x}_j$ for $i \neq j$ do not commute. Assuming 
$\Lambda$ as the dimensional scale for this commutation, it should be 
proportional to $\Lambda^{-2}$. Thus, it is sub-dominant with respect to 
(\ref{xppfcommu}) which is of order $\Lambda^{-1}$. In addition, one can 
assume that $f (\hat{x}_i, \hat{p}_i)$ depends also on $\hat{p}_j$. This does 
not change (\ref{xppijcommu}) but will add terms proportional to 
$\Lambda^{-1}$ to (\ref{xppicommu}). In this case $f (\hat{x}_i, \hat{p}_i, 
\hat{p}_j)$ is:
\be
f (\hat{x}_i, \hat{p}_i, \hat{p}_j) = \sum_{m,n,k} \frac{C_{mn} :\hat{x}_i^m 
\hat{p}_i^n \hat{p}_j^k:}{\Lambda^{n+k-m-1}} \label {fjexpan}
\ee
Assuming that $n$ and $k$ can take half-integer values, the term with 
$m = -1$ and $n = k = 1/2$ has the same order in $\Lambda$ as the term used to 
obtain equation (\ref{ppi}). The half-power of momentums can be related to a 
supersymmetric 
transformation and permits a natural extension of this model to supergravity 
and introduction of spinors in the model. We leave detailed investigation of 
this case to another work.

In the formulation above we neglected spacetime indexes. Apart from 
simplifying the notation, there is a deeper reason for this negligence. First 
of all it proves that this formalism can be applied to any background 
spacetime. More importantly, the similarity of the role of species/particles 
index to spacetime index indicates that there is an interchangeable role 
between what is called {\it a particle} and {\it a point in the spacetime}. 
Therefore we can claim that in this model there is a natural unification or 
embedding of the spacetime with a group manifold determining the symmetries 
and variety of particles. This aspect is similar to string theories. The hope 
is that in the extension of this model to continuum, geometrical properties 
and symmetries determine the signature of spacetime metric, the emergence 
of a time as well as the dimension of space.

\section{outline}
We suggested a modified structure of Hilbert space of multi-particle quantum 
system to present the effect of gravity. We showed that the redefinition of 
coordinate and momentum operators leads to a dynamical equation similar to the 
Wheeler-DeWitt equation for quantum mechanics in curved spacetime. A natural 
extension of the model to supergravity seems possible. Before being able to 
claim this model as a candidate model for quantum gravity we need to 
investigate a number of issues such as extension to infinite number of 
particles in another word a field theory, the emergence of a time coordinate, 
and possible relation between this model and other candidate quantum gravity 
models.

\begin{figure}[h]
\begin{center}
\includegraphics[width=10cm]{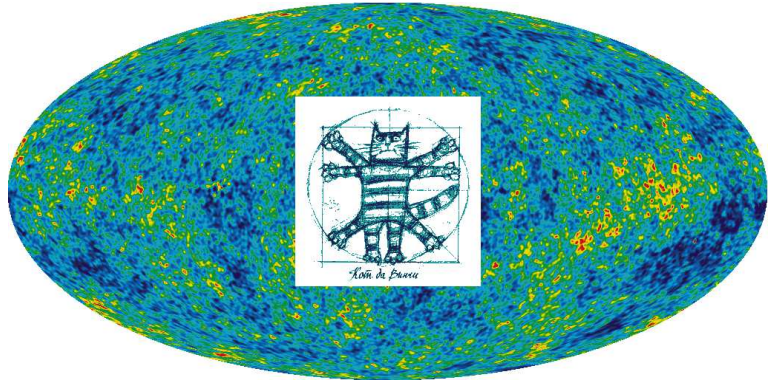}
\caption{\label{label} The Universe, presented in this figure by the map of 
Cosmic Microwave Background (CMB) anisotropy observed by the WMAP satellite, 
can be the large scale manifestation of an intrinsically indeterministic 
universe based on the uncertainty rules of Quantum Mechanics, symbolically 
presented by Schr\"odinger Cat \`a la Da Vinci to remind Tuscany, the home 
country of Da Vinci and the venue of the DICE2008 conference.}
\end{center}
\end{figure}

\subsection{Acknowledgments}
I would like to thank all the organizers of DICE2008 conference, notably Pr. 
Thomas Elze for their formidable organization and the choice of a wonderful 
place for the venue.

\end{document}